\documentclass[useAMS,usenatbib]{mnras}

\usepackage{graphicx}
\usepackage{scalefnt}
\usepackage{amssymb}
\usepackage{amsmath}
\usepackage{upgreek}
\usepackage{balance}

\newcommand{\yr}	    {\ifmmode \mathrm{yr} \else yr\fi}
\newcommand{\mpc}	{\ifmmode \,\mathrm{Mpc}^{-3} \else \,Mpc$^{-3}$\fi}
\newcommand{\Msun}	{\ifmmode \,\mathrm M_{\odot} \else $\,\mathrm M_{\odot}$\fi}
\newcommand{\Mhalo}	{\ifmmode M_{\mathrm{halo}} \else
  $M_{\mathrm{halo}}$\fi}
\newcommand{\Rvir}	{\ifmmode R_{\mathrm{crit,200}} \else $R_{\rm
    crit,200}$ \fi}
\newcommand{\Moff}	{\ifmmode M_{\mathrm{off}} \else
  $M_{\mathrm{off}}$\fi}
\newcommand{\Toff}	{\ifmmode T_{\mathrm{off}} \else $T_{\mathrm{off}}$\fi}
\newcommand{\Mstar}	{\ifmmode {M}_{\star} \else ${M}_{\star}$\fi}
\newcommand{\Mvir}	{\ifmmode M_{\mathrm{crit,200}} \else $M_{\rm
    crit,200}$ \fi}

\title[Metallicity-dependent IMF]{Simulating a metallicity-dependent
  initial mass function: Consequences for feedback and chemical
  abundances}
\author[T. A. Gutcke et al.]{Thales
  A. Gutcke$^1$\thanks{thales@mpia.de} and Volker Springel$^{2,3,4}$
\\
$^{1}$Max-Planck-Institut f\"ur Astronomie, K\"onigstuhl 17, 69117
Heidelberg, Germany\\
$^{2}$Heidelberg Institute for Theoretical Studies, Schloss-Wolfsbrunnenweg 35, 69118 Heidelberg, Germany\\
$^{3}$Zentrum für Astronomie der Universität Heidelberg, ARI, Mönchhofstrasse 12-14, D-69120 Heidelberg, Germany\\
$^{4}$Max-Planck-Institut für Astrophysik, Karl-Schwarzschild-Str. 1, D-85748, Garching, Germany\\}
\begin{document}
\pagerange{\pageref{firstpage}--\pageref{lastpage}} \pubyear{---}
\maketitle
\label{firstpage}
%==========================================================================%
\begin{abstract}
% <=250 words, usually no references
% Cosmological, hydrodynamical simulations commonly assume a
% universal stellar initial mass function (IMF)
%  when attempting to
% produce realistic Milky Way analogues.
  Observational and theoretical arguments increasingly suggest that
  the initial mass function (IMF) of stars may depend systematically
  on environment, yet most galaxy formation models to date assume a
  universal IMF. Here we investigate simulations of the formation of
  Milky Way analogues run {\itshape ab initio} with an empirically derived
  metallicity-dependent IMF and the moving-mesh code {\small AREPO} in
  order to characterize the associated uncertainties.  In particular,
  we compare a constant Chabrier and a varying metallicity-dependent
  IMF in cosmological, magneto-hydrodynamical zoom-in simulations of
  Milky Way-sized halos. We find that the non-linear effects due to
  IMF variations typically have a limited impact on the morphology and the star formation histories of the formed
  galaxies. Our results support the view that constraints on
  stellar-to-halo mass ratios, feedback strength, metallicity
  evolution and metallicity distributions are in part degenerate with
  the effects of a non-universal, metallicity-dependent
  IMF. Interestingly, the empirical relation we use between metallicity
  and the high mass slope of the IMF does not aid in the quenching
  process. It actually produces up to a factor of 2-3 more stellar
  mass if feedback is kept constant.  Additionally, the enrichment
  history and the $z=0$ metallicity distribution are significantly
  affected. In particular, the alpha enhancement pattern shows a
  steeper dependence on iron abundance in the metallicity-dependent
  model, in better agreement with observational constraints.
\end{abstract}
%==========================================================================%
\section{Introduction}
%
%==========================================================================%
%
The majority of stars are known to form in shared star formation (SF)
regions known as embedded clusters \citep{LadaLada2003, Kroupa2005,
  Megeath2016}. The cause of this is assumed to be the collapse of
giant molecular clouds (GMC) where many stars form
simultaneously. This implies a distribution of stellar masses
originating from each collapse, the stellar initial mass function
(IMF). An unanswered question in stellar and galaxy formation physics
is whether the IMF is universal, i.e.~whether the distribution is the
same in each star forming region or varies systematically with some
dependency on its environment.

% Theoretically, a comprehensive physical picture explaining the origin and properties of the IMF does not exist yet; to this regard, Silk (1995); Krumholz (2011) analyzed the effect of molecular flows and protostellar winds, \citet{Larson1998, Larson2005} tried to explain it in terms of the Jeans mass, while Bonnell et al. (2007); Hopkins (2013); Chabrier et al. (2014) explored the effect of gravitational fragmentation and of the thermal physics.

The details of the physics governing the regulation of star formation
are poorly understood, since it is still unclear to what degree
radiation, turbulence, magnetic fields, and instantaneous feedback
from the forming young stars are able to counteract a pure
gravitational collapse model. But observations of the Milky Way (MW)
Galaxy, including resolved imaging of individual stars in globular
clusters and the use of dwarf and giant sensitive spectral features,
are broadly in agreement with a universal IMF shape in the MW
\citep{Scalo1986, Kroupa2001, Chabrier2003,
  Bastian2010,Hopkins2013}. Pioneering observational work on the IMF
and a first functional fitting form were made by
\citealt{Salpeter1955}, who suggested a power law slope of
$1.35$. Later, \citet{Kroupa2001} and \citet{Chabrier2003} added a
low mass turn over to this basic model.

Studies of other galaxies yielded varying results. Some are consistent
with the results from the MW \citep{Kroupa2002, Chabrier2003,
  KirkMyers2011, Bastian2011}. {But \citet{Larson1998} showed that most
observations are compatible with a moderate IMF variation in
time. For example, both the G-dwarf problem and the large amount of heavy elements
in clusters can be reconciled with a change by a factor of $\sim3$ in
the dwarf-to-giant ratio.}
Recently, further evidence for IMF
variations has emerged, especially in early-type galaxies
\citep[e.g.][]{HoverstenGlazebrook2008, Meurer2009, Gunawardhana2011, Conroy2012,
  Cappellari2013}.  Studying resolved star clusters in M31, \citealt{Weisz2015} found a steeper
high-mass power law slope of $(1.45\pm0.05)$. \citet{MartinNavarro2015} used a sample of CALIFA
elliptical galaxies to demonstrate a dependency of the IMF on the
metallicity of the gas from which the stars formed. They use the
dwarf-to-giant ratio and allowed the high-mass slope,
$\Gamma_{\rm b}$, of the Vaszdekis IMF \citep{Vazdekis1996} to vary.

{Since compact ETGs are believed to be
stellar dominated at their centers, one way of measuring the IMF of
integrated light is to use the mass-to-light ratio as a proxy for
the fraction of low-mass stars \citep[e.g.][]{Dutton2011}. Using this method, \citet{Conroy2013} analyze a sample of
compact elliptical galaxies and find IMF variations with velocity
dispersion ($\sigma$). They
show that galaxies with $\sigma \sim 100$ km s$^{-1}$ have IMF slopes most
similar to the Milky
Way. Intermediate velocity dispersion galaxies ($\sigma \sim
160$ km s$^{-1}$) become more bottom-heavy, with an IMF shape best fit by a
Salpeter IMF. The highest velocity dispersion galaxies with $\sigma \sim 250 -
300$ km s$^{-1}$ have yet more bottom-heavy IMFs.}
\citet{LaBarbera2013} attempt to put constraints on the IMF using a
large spectroscopic sample of early-type galaxies from the SPIDER
survey. They also find a correlation
between the IMF slope $\Gamma$ and the central velocity dispersion. A
similar result is obtained by \citet{Spiniello2014}, who use a set of galaxies from the MILES
library of stellar spectra to show a dependency of the IMF slope on
the global velocity dispersion.

\citet{Chabrier2014} examine the evidence in the literature supporting
a variable IMF in ``extreme starburst environments'' and in massive
early-type galaxies (ETGs). They show that ETGs sport spectral
absorption features, larger mass-to-light ratios and larger
[$\alpha$/Fe] than spiral galaxies. ETGs therefore are assumed to host
more low-mass stars such as M dwarfs. The authors postulate that these
local ETGs evolved from small, compact dense progenitors between
$z\sim 3-5$. These very compact systems were $10^3$ times denser than
the Milky Way (MW), had higher velocity dispersions and possibly
higher temperatures. Also, these progenitors had different turbulent
velocity than the MW, producing intense pressure and high accretion
rates. The IMF in these ``extreme starburst environments'' is thus
expected to be bottom-heavy with respect to the MW.

Simulations have also been used to study the effect of an IMF
variation on the properties of galaxies. \citet{Blancato2017}
post-process the Illustris simulation to study the effects of an IMF
dependency on local and global velocity dispersion and on the
metallicity, finding that the hierarchical nature of massive galaxy
assembly significantly complicates the interpretation of the observed
trends.

\citet{Yan2017} build on the integrated galactic IMF (IGIMF) model
\citep{Kroupa2003}. This model defines the galaxy-wide IMF as the
integration of both the local IMF of a single embedded star cluster
and the cluster mass function. Smaller clusters can be ``optimally
sampled'' from a universal local IMF, but given their total mass, will
not produce any stars more massive than the cluster mass. The result
for galaxies as a whole is that a universal local IMF can produce a
IGIMF that varies.

\citet{Clauwens2016} also use the metallicity-dependent IMF from
\citet{MartinNavarro2015}, applying it to a sample of SDSS galaxies to
investigate the effect on total stellar mass and star formation
rates. They find that a bottom-heavy IMF at late times can aid in the
quenching process, speeding it up.  \citet{Guszejnov2017} use the FIRE
simulation suite to study the IMF variation caused by a number of
different cloud collapse models, and show that only a protostellar
feedback model produces small enough variations to be consistent with
observations.

This paper is organised as follows. Section~\ref{sec:sims} recaps the
simulation code and physics, Section~\ref{sec:model} describes the
variable IMF model, Section~\ref{sec:obs} compares the Chabrier and
variable IMF results, and Section~\ref{sec:summary} summarises our
findings. We use the cosmological parameters from the most recent
Planck results \citep[$\Omega_{\rm m} = 0.31$,
$\Omega_{\rm L} = 0.69$, $\Omega_{\rm b} = 0.0486$, $h = 0.677$,
$\sigma_{8} = 0.8159$, $n_s = 0.97$][]{Planck2016}. We assume solar
metallicity, $Z_{\odot}$, to be $0.0127$. We use the variables \Rvir
and $R_{\rm vir}$ interchangeably to denote the radius inside which the
average density is 200 times the critical density.
%
%
%==========================================================================%
\section{Numerical Simulations}
\label{sec:sims}
%==========================================================================%
%
We have run six zoom-in simulations of Milky Way mass galaxies
($\sim10^{12}\Mvir$ at $z=0$).  The initial conditions are drawn from
the Auriga zoom-in simulation project presented in \citet{Grand2017a},
and correspond to their galaxy numbers 6, 9, 13, 16, 24, and 28.  We
use the magneto-hydrodynamical moving-mesh code {\small AREPO}
\citep{Springel2010} with a physics setting that differs slightly from
the Auriga project with respect to parameters used for black hole
seeding, active galactic nuclei feedback, and galactic winds. In
articular, we use here the updated yield tables introduced in the
IllustrisTNG simulation \citep{Pillepich2017,Naiman2017} project.
% The chemical tagging that differentiates the
% production of metals from SNIa, SNII and AGB stars is described in
% \citep{Naiman2017}.
These changes in the details of the physics
implementation with respect to Auriga are however immaterial for the
qualitative results of our simulations. Nevertheless, to differentiate
the true Auriga galaxies from our re-simulations also in the universal
IMF case (which have slightly different physics), we will henceforth
call our galaxies ``Halo-X'', where `X' denotes the corresponding
number 6, 9, 13, 16, 24 and 28 in the Auriga project.

The IMF slope effects both the SNII rate and the AGB wind metal
enrichment. The SNII rate at a given timestep is determined by
\begin{equation}
N_{\rm SNII}(t, \Delta t) = \int_{M_{\rm low}}^{M_{\rm high}}{\Phi(m)\, {\rm d}m}.
\end{equation}
For SNII, $M_{\rm low} = {\rm max}[M(t+\Delta t), 8\Msun]$ and $M_{\rm high} =
{\rm min}[M(t), 100\Msun]$, where $M(t, Z)$ is the inverted stellar lifetime
function \citep[see][section 2.3.1]{Vogelsberger2013}. The yields for
AGB stars are taken from \citet{Karakas2010} for $1-6\Msun$, and from
\citet{Doherty2014} and \citet{Fishlock2014} for $7-7.5\Msun$. The
SNII yields are taken from \citet{Portinari1998} for $6-13\Msun$ and
$40-120\Msun$. For $13-40\Msun$, they are from \citet{Kobayashi2006}.

The mass fraction of SNIa (in the range 1.4-8~\Msun) for different IMF
is not changed in our model. Likewise, the ``delay time distribution''
function, $g(t)$, is taken in unmodified form from the
\citet{Pillepich2017} model. {Since the binary fraction and the
  time delay distribution of SNIa are very uncertain, we did not attempt to couple them to the
variable IMF model.}
{We make this choice for consistency, because the fiducial model assumes them to be fixed and independent of the IMF.

The SNIa rate is nevertheless indirectly affected by the changed star
 formation history. It is computed as}
\begin{equation}
\dot{N}_{Ia}(t) = \int_0^t{\Psi(t')g(t-t')\,\text{d}t'} ,
\end{equation}
where $\Psi(t')$ is the star formation rate and $g(t-t')$ is the delay
time distribution (DTD) defined as
\begin{equation}
g(t) = 
\begin{cases}
0 & \text{if $t<40$Myr}\\
N_0 \Bigg( \dfrac{t}{40\text{Myr}}\Bigg)^{-s} \dfrac{s-1}{40\text{Myr},} & \text{if $t \geqslant40$Myr,}
\end{cases}
\end{equation}

The SNIa yields are taken from \citet{Nomoto1997}.

The fiducial resolution level of the bulk of our simulations
corresponds to ``level 5'' as tabulated in \citet{Grand2017a}. We also
run one ``level 4'' version of Auriga halo number 16 as a resolution
test. The fiducial baseline IMF when universality is assumed is the
Chabrier IMF \citep{Chabrier2003}, which we compare with a simple
metallicity-dependent IMF model described in the following section.
%
%==========================================================================%
\section{The variable IMF model}
\label{sec:model}
%==========================================================================%
We use the relation derived in the study by \citet{MartinNavarro2015}
as the starting point of our simulations. {The relation
  presented in \citet{MartinNavarro2015} is interesting to explore because
  it is a recent study and because metallicity is a very plausible
candidate for IMF variations from a theoretical perspective. Since
this study is intended as an
initial exploration, we leave the other empirical relations presented in the literature
\citep[e.g.][]{HoverstenGlazebrook2008, Meurer2009, Gunawardhana2011, Marks2012} for future work.}

The empirically derived
relation between the high-mass slope of a Vasdekis IMF,
$\Gamma_{\rm b}$, and the [M/H] of the gas in \citet{MartinNavarro2015} is:
\begin{equation}
\Gamma_{\rm b} = 2.2(\pm0.1) + 3.1(\pm0.5) \times {\rm[M/H]}.
\label{eq:MartinNavarro}
\end{equation}
\begin{figure}
\centering
\includegraphics[width=0.5\textwidth]{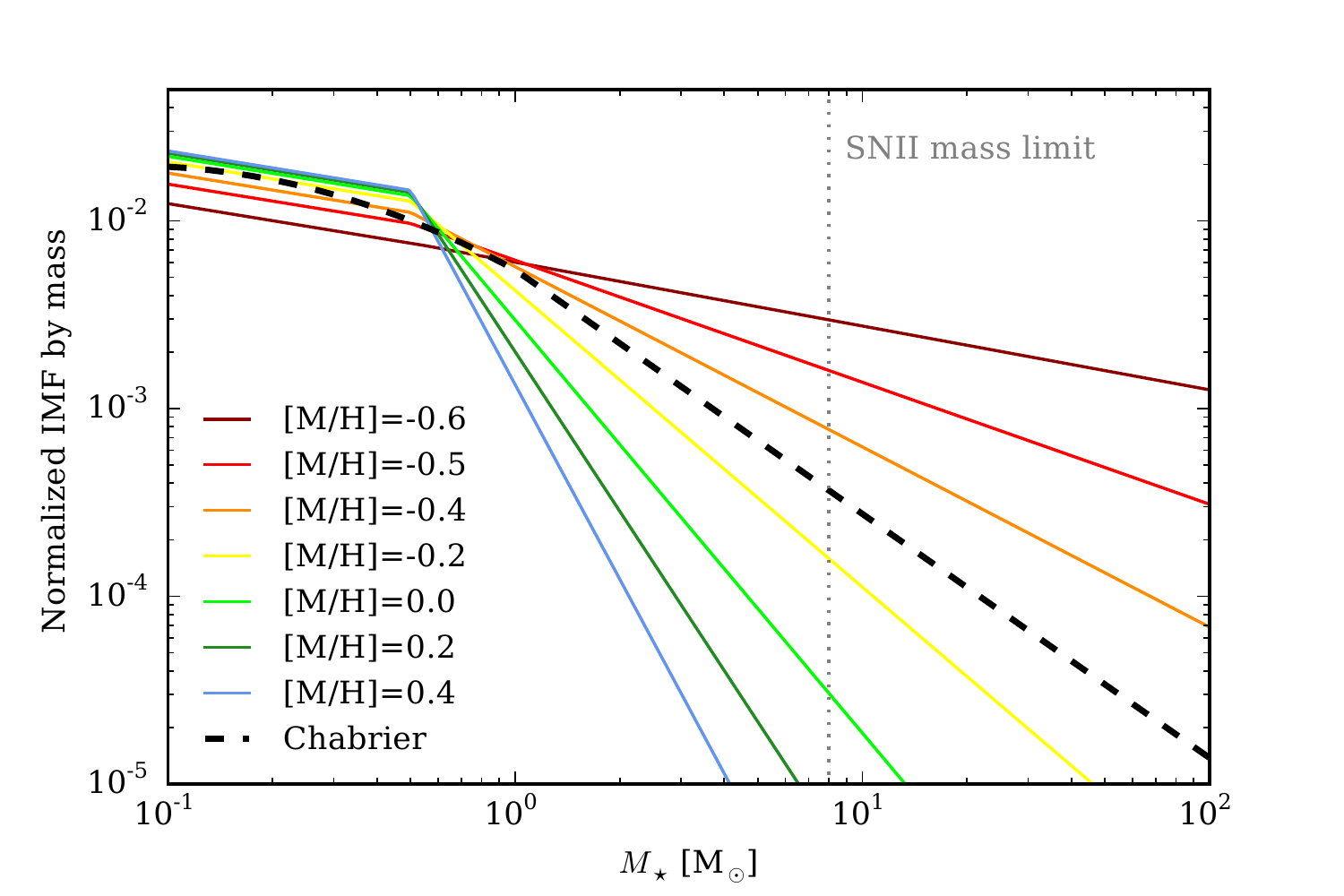}
  \caption{Illustration of the varying IMF (by mass) with different [M/H]
    values. The dashed black line is the Chabrier IMF. The grey dotted
  line indicates the lower mass limit for core collapse supernovae.}
  \label{fig:imf}
\end{figure}
We use this relation, but combine it with the Kroupa IMF
\citep{Kroupa2001} to define a metallicity dependent IMF as follows:
\begin{equation}
\Phi(m)\propto
\begin{cases}
m^{-0.3} & \text{for $0.1 <m/\Msun<0.5$}\\
m^{-\Gamma_{\rm b}} & \text{for $0.5 <m/\Msun<100$}.
\end{cases}
\end{equation}
The integral of the IMF is normalised to the total initial stellar mass of a
star particle, $M_{\rm init}$. This mass has a resolution-dependent
target value corresponding to our gas-cell resolution but can
vary slightly around the target value as described in
\citet{Vogelsberger2013}.
We extract the value for [M/H] from the simulation as
\begin{equation}
{\rm[M/H] = log}_{10}\Bigg(\frac{\sum{\frac{M_{\rm Z}}{M_{\rm tot}}}}{Z_{\odot}}\Bigg) ,
\end{equation}
summing over all $Z$ above He.

In the following, we will use the terms ``bottom-heavy'' and
``top-heavy'' to refer to IMFs with large and small (or even negative)
$\Gamma_{\rm b}$, respectively. Given that each IMF is re-normalized
to $M_{\rm init}$, a steep high-mass slope will leave more stellar
mass in the low-mass region. Thus, these terms only refer to the
relative amount of $M_{\rm init}$ in the low-mass ($\lesssim 1\Msun$)
and high-mass ($\gtrsim 1\Msun$) regions of the IMF.

Since the observations of IMF variations are as of yet inconclusive,
the common choice made by most galaxy formation simulations is to
assume universality (an Occam's razor argument).  The primary goal of
our present analysis is to assess the magnitude of potential changes
in global galaxy properties that a variable IMF consistent with
observations can produce. This then provides an approximate
determination of the uncertainties associated with making the ad-hoc
choice of a universal IMF.

Most simulations studying variable IMF models use simulations run with
a constant IMF and post-process these to re-calculate galaxy
properties assuming a variable IMF. These models cannot easily account
for the non-linear effects caused by variability, i.e.~for the changes
in star formation rates and associated feedback resulting from
variations in the enrichment history due to a different IMF.  Thus, a
second goal of our analysis is to understand such non-linear effects
on the metallicity that a variable IMF can cause. Hence, we will
investigate the enrichment history and the metallicity distribution
functions (MDFs) of our simulated galaxies.

We note that \citet{MartinNavarro2015} used early type galaxies to
derive their empirical relation, whereas our study focuses on Milky
Way-like spiral galaxies. While there is no direct evidence that the
relation also applies to late type galaxies, we here shall assume that
it does. {We make this simplifying assumption mainly because the
  Milky Way is the most important Galaxy for chemo-dynamical
  studies. Examining other galaxy mass regimes is beyond the scope of
  this paper.} Conceptually, this also means that only one new parameter (in our
case the metallicity of the star forming gas) is added when moving
from the assumption of a fixed universal IMF to a universally varying
IMF. This thus represents the next simplest model in which the
variation of the IMF with metallicity is identical in all types of
galaxies.

%
%==========================================================================%
\section{Results}
\label{sec:obs}
%==========================================================================%
%
\subsection{Stellar mass evolution}
%==========================================================================%
%
\begin{figure}
\centering
\includegraphics[width=0.2\textwidth]{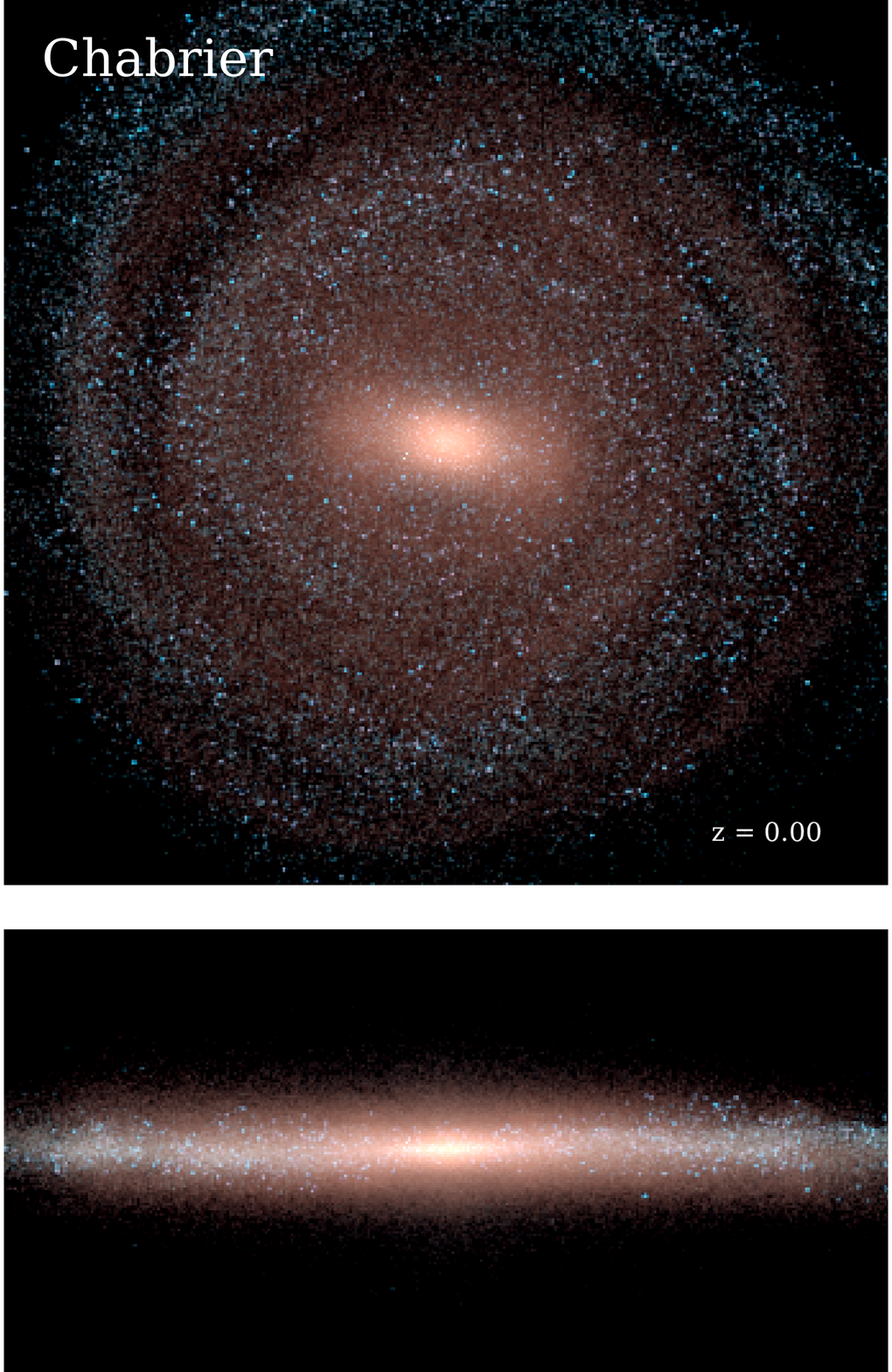}
  \includegraphics[width=0.2\textwidth]{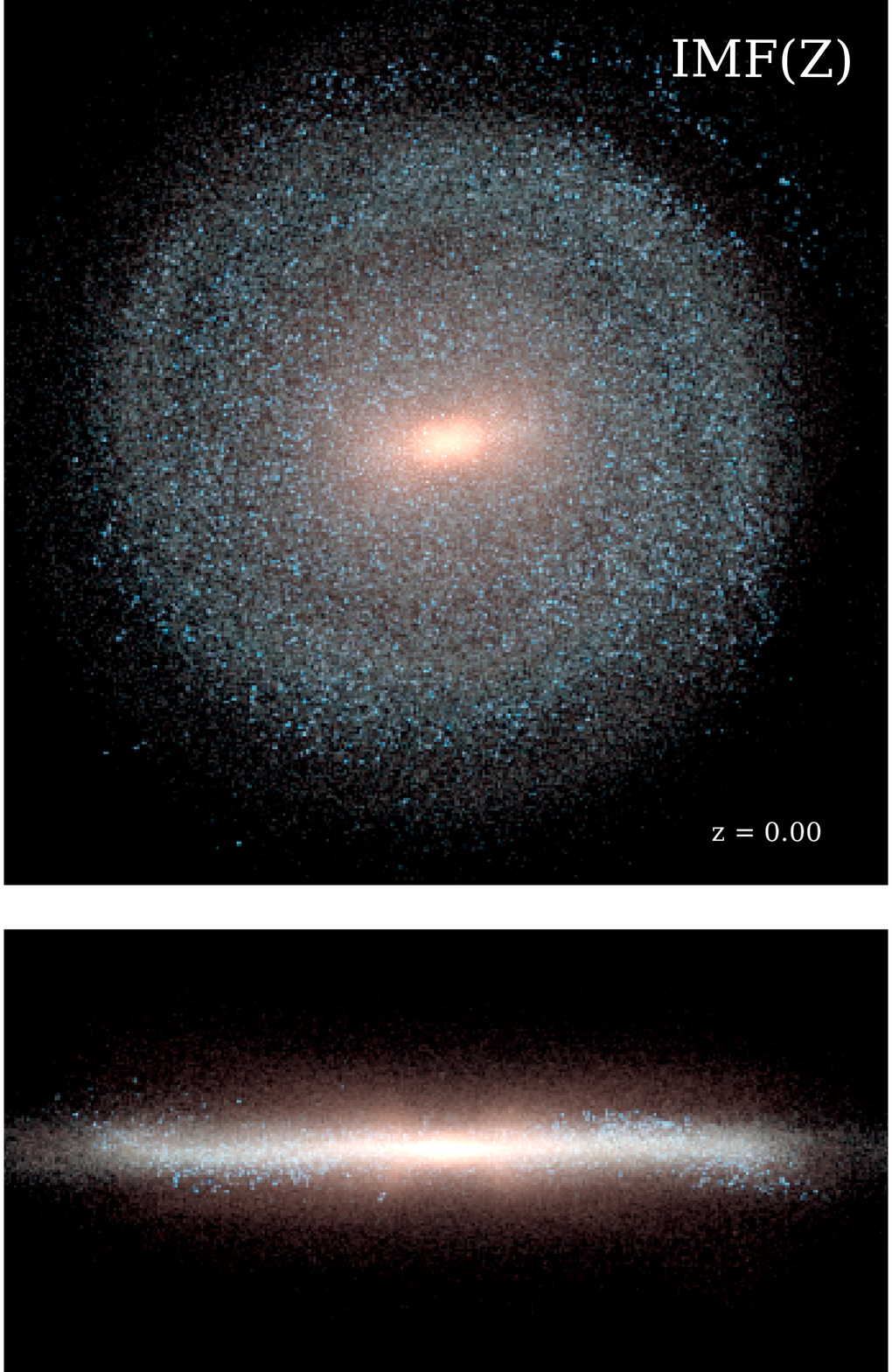}
  \caption{RGB images of Halo 16 for Chabrier IMF (left) and the
    metallicity dependent IMF (right). The galaxy is rotated to see
    the disk face-on in the upper panels, and edge-on in the lower
    panels. The images are 50\,kpc wide. The K-, B-, and U-filters
    were mapped to RGB channels, respectively. We note that the
    photometric calculations assume a Chabrier IMF in both cases.}
  \label{fig:images}
\end{figure}
\begin{figure*}
  \centering
\includegraphics[width=0.48\textwidth]{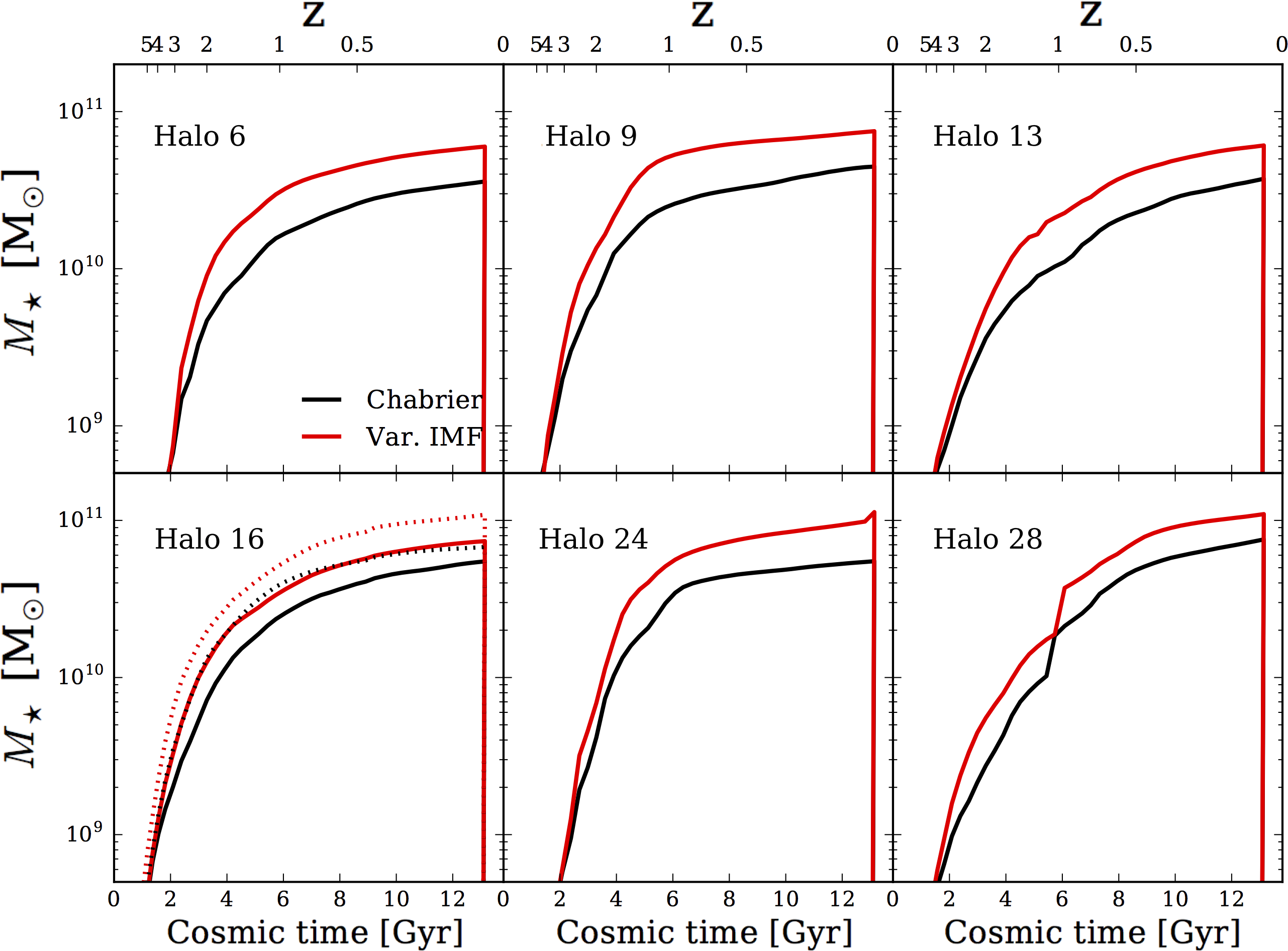}
\hspace{0.02\textwidth}
\vspace{0.04\textwidth}
 \includegraphics[width=0.48\textwidth]{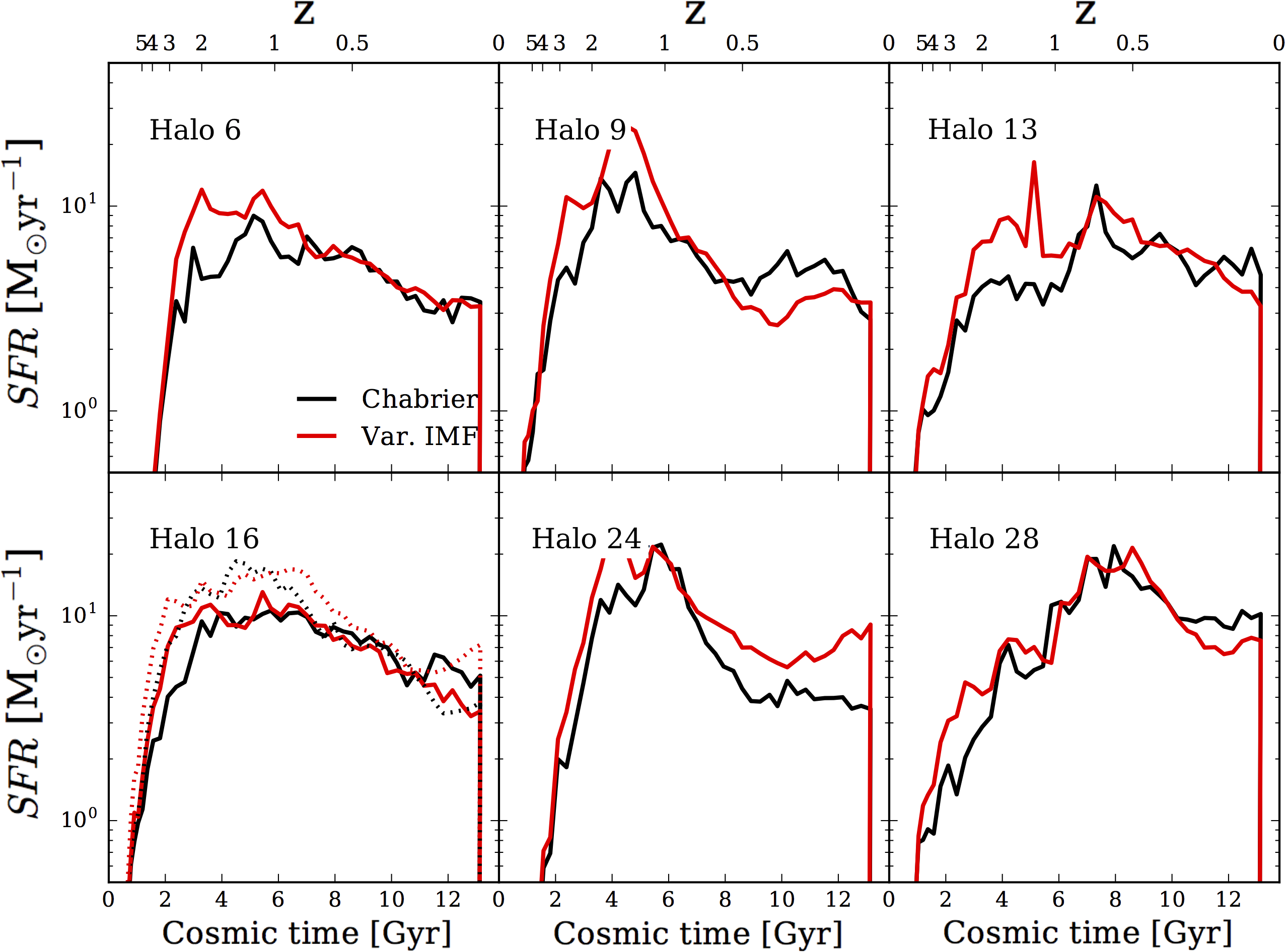} 
  \includegraphics[width=0.48\textwidth]{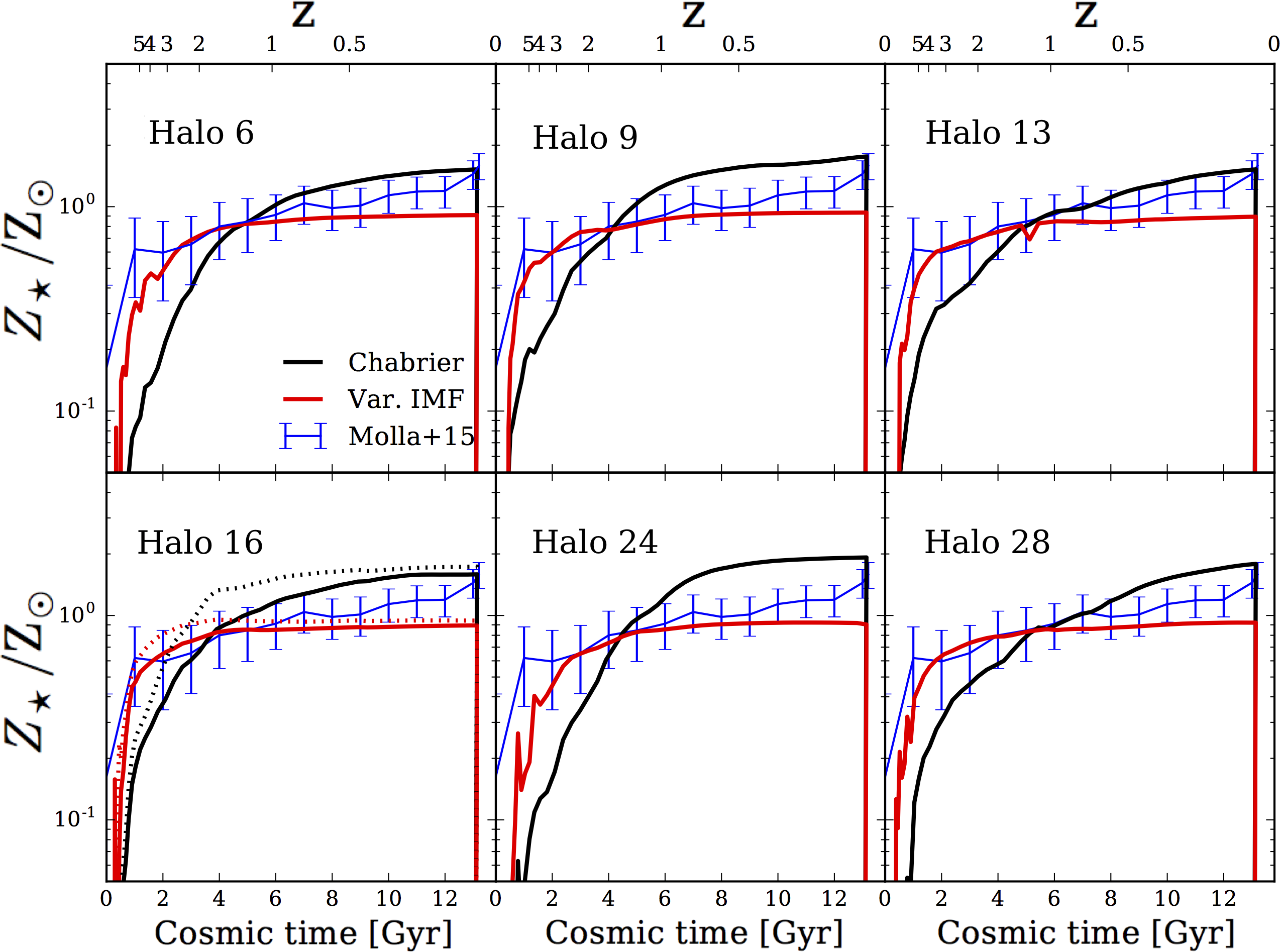}
\hspace{0.02\textwidth}
\includegraphics[width=0.48\textwidth]{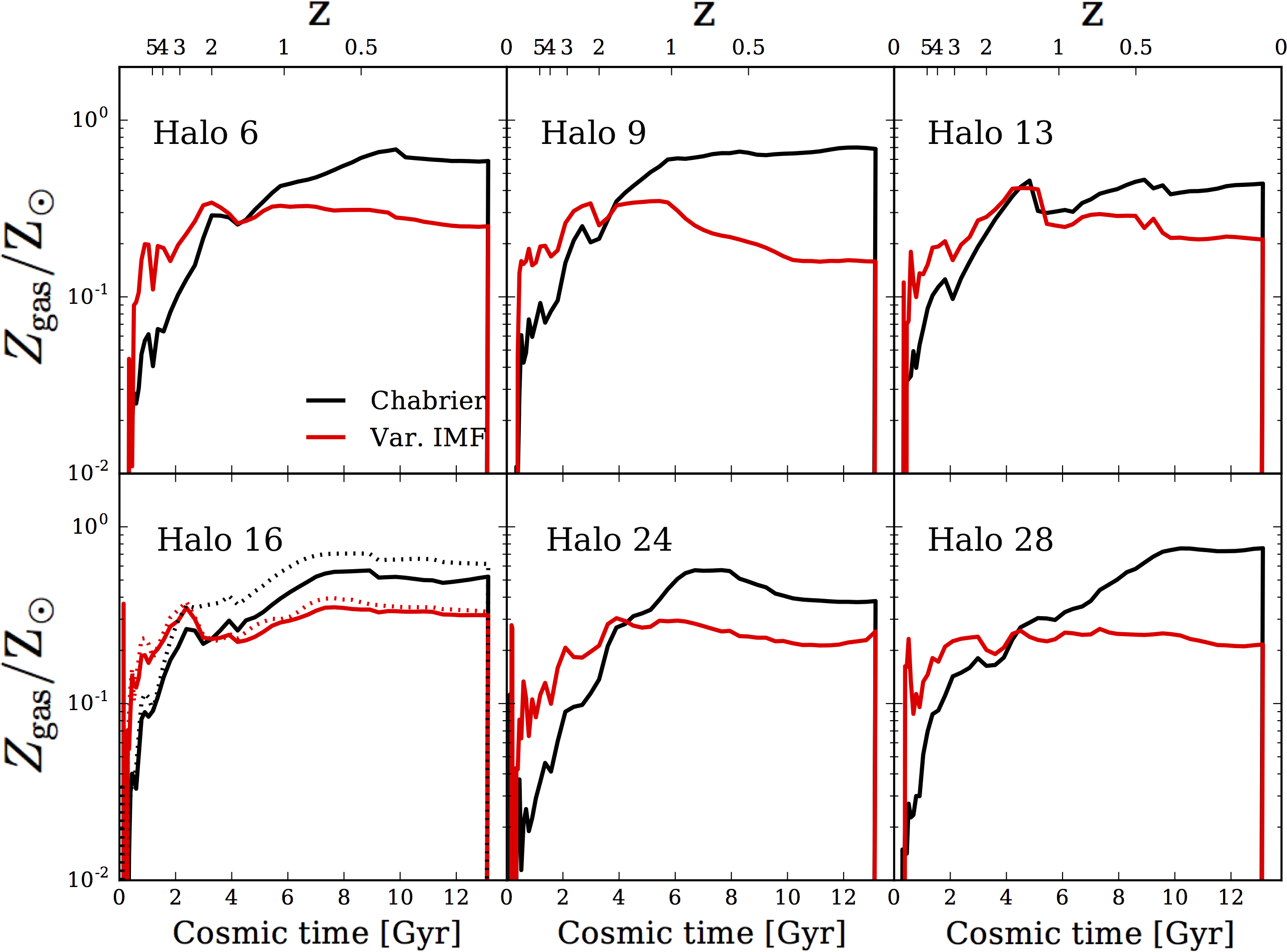}
\caption{ {\itshape Upper panels:} Stellar mass evolution (left) and
  star formation histories (right) for the six halos (initial
  conditions from the Auriga galaxies 6, 9, 13, 16, 24, 28). The black
  lines are the galaxies assuming a universal Chabrier IMF. The red
  lines are the same galaxies re-simulated with the
  metallicity-dependent IMF. The stellar mass is the sum of all star
  particle masses inside \Rvir at each timestep. The dotted lines in
  the panel with Halo 16 are the higher resolution versions of the
  same initial conditions. {\itshape Lower panels:} Total stellar
  (left) and gas (right) metallicity evolution of all stars within
  \Rvir for each galaxy. Lines are the same as in
  Figure~\ref{fig:starmassevolution}. The blue error bars are the data
  points from \citet{Molla2015}, who compiled a variety observational
  studies of the Solar neighbourhood.}
  \label{fig:sfh}
\label{fig:starmassevolution}
  \label{fig:Zstar}
\label{fig:Zgas}
\end{figure*}

Given the mutual self-regulated coupling of star formation and stellar
feedback it is difficult to predict a priori whether our variable IMF
simulations would produce more or less stellar mass than their
Chabrier counterparts. Since metallicity is low at early times, the
IMF in our model has a shallow slope then (being ``top-heavy''),
allowing for many high mass stars. These in turn produce a lot of
core-collapse supernovae (SNII) and stellar feedback. The heating from
the feedback can inhibit further star formation. On the other hand,
the increase in high mass stars at early times also produces
significant early enrichment in the gas. This enrichment steepens the IMF slope of
subsequent generations of stars, thus decreasing the amount of
feedback energy effecting the gas in subsequent cycles.  A critical question
then is, how long does it take to enrich the gas sufficiently that the IMF
is truncated below $8 \Msun$, the minimum mass of core-collapse
supernovae. Once this occurs, both chemical enrichment and stellar
feedback come largely to a halt.  Since our IMF variations do not
affect the total mass of a formed star particle but only the
distribution of its mass (the mass function), stellar mass build up
can then proceed without the effect of feedback, due to the lack of
high mass stars, and is solely regulated by the available gas
reservoir after this point.

A bottom-heavy IMF and the resulting lack of stellar feedback has the
further effect that less mass is returned to the interstellar medium
(ISM) and becomes again available for subsequent star formation.
Depending on the precise dependency of the IMF slope with metallicity,
and on the efficiency of stellar feedback, these two processes can
dominate at different times in a galaxy's life.

Figure~\ref{fig:images} shows $z=0$ RGB images of Halo 16 in face-on
and edge-on projections for both our models. The RGB channels use the
K-, B- and U-filters based on the stellar population synthesis model
by \citet{BruzualCharlot2003}. We note however that the image was
created with the standard photometric calculation that assumes a
Chabrier IMF, for simplicity. Thus, young stars (blue) are a bit
brighter than they would self-consistently appear if the variable IMF
was also used for making the image, since the variable model produces
a bottom-heavy IMF at late times.

Figure~\ref{fig:starmassevolution} gives an overview of the stellar
mass evolution of our six halos. The black solid line is the evolution
with the universal Chabrier IMF, while the red line is the variable
IMF model. The dotted lines show Halo 16 at resolution level 4. In all
cases, the variable IMF model produces more stellar mass at early
times ($z\sim 2-3$) and keeps the lead for the entire evolution. The
final stellar mass is a factor of 2-3 higher for the variable model in
all model galaxies.

We stress that the feedback model was calibrated for the universal IMF
and no adjustments were made to account for the variable IMF
scenario. {Thus, we do not claim that the variable IMF model alone
  can explain the observations. Instead,} the change in final stellar mass should only
be interpreted as a relative change from model to model when only the
IMF parameterisation is changed, but other aspects of the model are
left unchanged. Of course, if the variable IMF model was adopted as a
new default description of the galaxy formation physics, the feedback
strength would have to be re-calibrated. Increasing the feedback would
likely again be able to match the stellar mass -- halo mass relation.

Figure~\ref{fig:sfh} shows the star formation histories of our six
halos. The star formation rates are higher at early times in the
variable IMF case, but drop below the Chabrier model in all but one of
our halos by $z=0$. Due to the shortage of core-collapse SN after
$z\sim3$, the gas in the galaxy becomes calm and is not driven into
outflows. Thus, it is consumed by SF quickly at that time. The stars
formed have a bottom heavy IMF that stores much of the mass for times
close to the Hubble time. Thus, SF then decreases, in most cases below
the Chabrier level.

%
%==========================================================================%
\subsection{Metallicity evolution}
\label{sec:metals}
%==========================================================================%
Stellar metallicity is what determines the IMF and subsequent
evolution of each star particle. This is where the two IMF models show
their most significant differences.  The evolution of the total
stellar metallicity of each halo is shown in
Figure~\ref{fig:Zstar}. The blue line and error bars are the binned MW
data compilation from \citet{Molla2015}. All six halos in the variable
model increase their global metallicity dramatically in the first
2\,Gyr. After that, the evolution flattens out and stays practically
constant for the last 8\,Gyr till $z=0$. The compiled observations
from \citet{Molla2015} allow for such a rapid increase at early
times. But by $z=0$, the Chabrier value better matches the data.

The gas phase metallicity evolution across the entire halo is shown in
Figure~\ref{fig:Zgas} and sports the same change from Chabrier IMF to
variable model. The total gas metallicity increases faster and earlier
in the metallicity dependent case, and then slows its evolution, until
no significant increase in total metallcity happens in the last
7-8\,Gyr. 

This similar trend in both the stellar and gas metallicity in the
variable model can be explained by the change in IMF slope across
time. Initially, the gas has a primordial composition, with no
metals. Thus, the IMF allows for many high-mass stars that undergo
core collapse within the first 1-2\,Gyr. This increases the
metallicity rapidly. After this first phase, the gas has been enriched
and the subsequent generation of stars have IMFs with much steeper
slopes and fewer high-mass stars. The strong enrichment stops and the
metallcity of both gas and stars does not increase much further.

\begin{figure}
  \centering
 \includegraphics[width=0.5\textwidth]{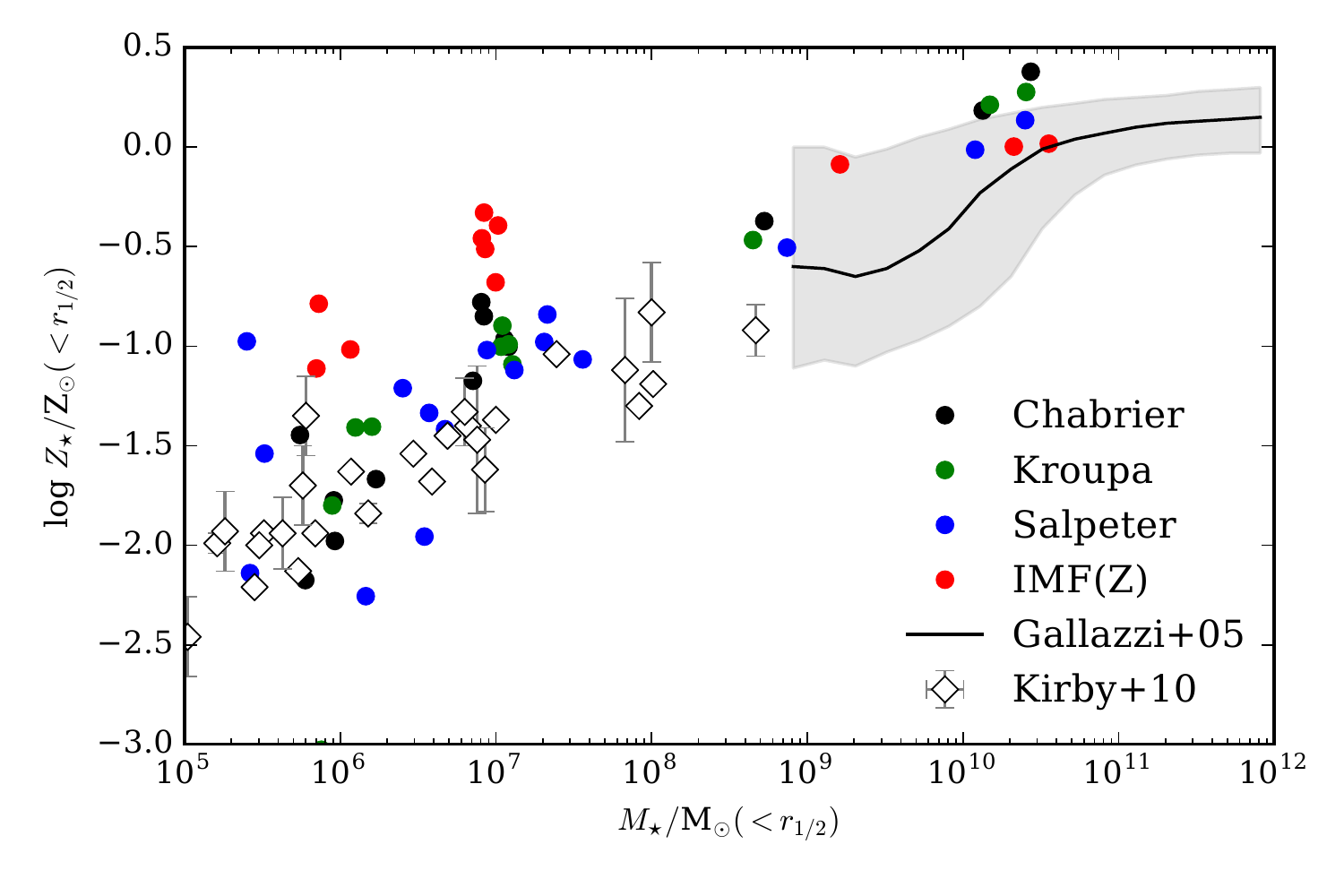}
 \caption{Stellar metallicity within the half mass radius, $r_{1/2}$, as
   a function of stellar mass. Dots represent the ten most massive
   halos/subhalos in the zoom simulation Halo 16. Three universal IMFs
   (Chabrier, Kroupa and Salpeter) and the metallicity dependent IMF
   are shown. The gray band shows the SDSS data from
   \citet{Gallazzi2005}. Gray data with error bars show the MW dwarfs from \citet{Kirby2010}.}
  \label{fig:gallazzi}
\end{figure}

In Figure~\ref{fig:gallazzi}, we compare our galaxy stellar
metallicity to the observational data from \citet{Gallazzi2005} who
used SDSS data. We match the observational selection by cutting the
view at 5\,kpc, which is the average physical size of the galaxies
inside the SDSS fiber. The points show the ten most massive (and
resolved) galaxies/satellites for each simulation. We compare
Chabrier, Kroupa, Salpeter and variable IMF models. At low masses, we
have added the MW satellite data from \citet{Kirby2010}. At these low
masses ($\Mvir<10^{8}\Msun$), the universal IMF models match the data
better. Interestingly, at higher masses (the highest being the MW-like
galaxy itself), the results for the metallicity-dependent IMF are
closer to the mean of the \citet{Gallazzi2005} data.
%
%==========================================================================%
\subsection{Metallicity distribution functions}
\label{sec:mdf}
\begin{figure} 
  \centering
 \includegraphics[width=0.5\textwidth]{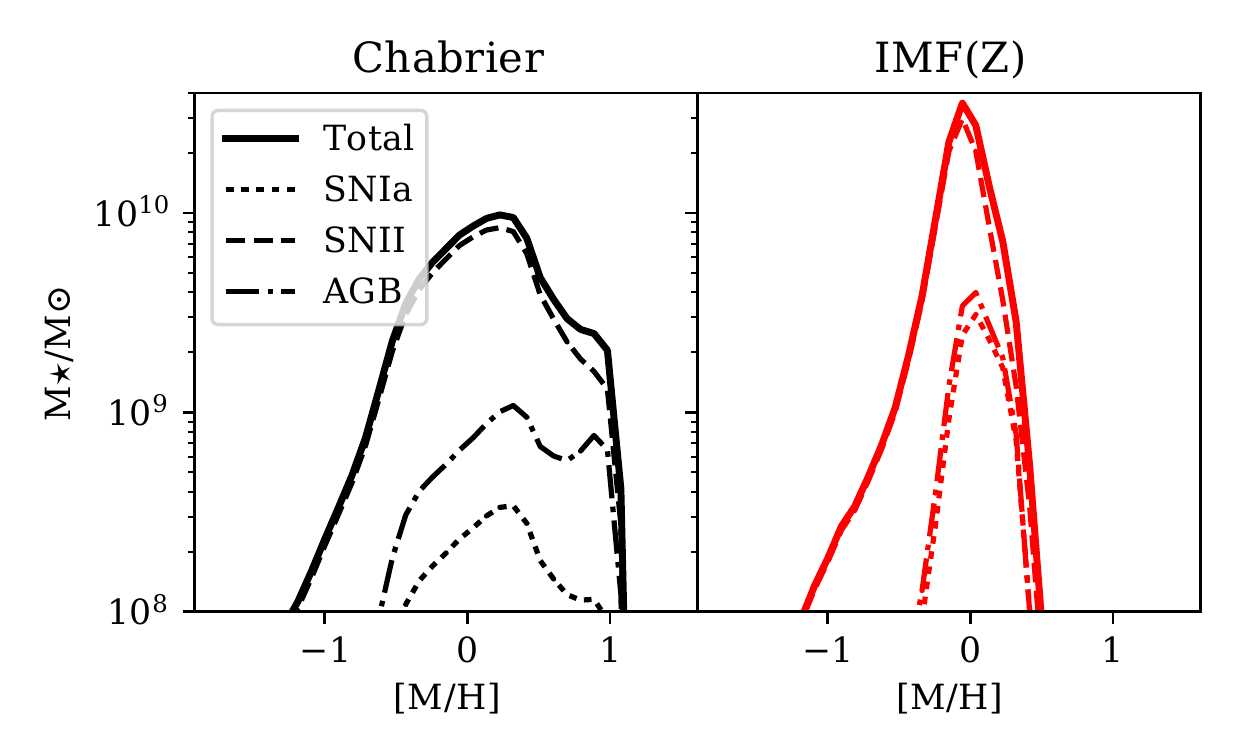}
 \caption{Stellar metallicity distribution function (solid line),
   split into contributions from SNIa (dotted), SNII (dashed) and AGB
   stars (dot-dashed). Shown is the $z=0$ simulation output of Halo 16 (level
   4).}
  \label{fig:stellarMDFtagged}
\end{figure}

%==========================================================================%
The metallicity distribution functions (MDFs) of the stars and gas are
a good way to track the changes caused by the variable IMF model,
since the metallicity dependence causes non-linear effects in the
metal production.  The stellar MDFs are shown in
Figure~\ref{fig:stellarMDFtagged}. Both models produce distributions
that peak around [M/H]$\simeq 0.0$, but the variable IMF model is
distributed less towards high metallicity. This is caused by the lack
of high-mass stars at late times, since the IMF drops off steeply. In
both cases the majority of metals are produced in core-collapse
supernovae (SNII, dashed lines). In the Chabrier case, asymptotic
giant branch stars (AGB) produce most of the remaining metals whereas
SNIa contibute very little. In the variable IMF model, AGB and SNIa
produce approximately the same amount.
%
%
% In figure \ref{fig:stellaroverFe}, we take a look at the abundance
% distributions of C, N, O, Ne, Mg and Si. We show only stars within
% $0.1\,\Rvir$. The standard $\alpha$
% elements O, Ne, Mg and Si all show a similar difference between the
% two models. The variable model produces a flatter, broader
% distribution. C and N peak at lower values and have a tail towards low
% abundances compared to the Chabrier case.
%
% We also take a look at the $\alpha$-abundance distributions in
% 0.1\,\Rvir and in the half-mass radius, $R_{0.5}$. We define $\alpha$
% as (O + Mg + Si)/3. The $\alpha$-abundance has been shown to correlate
% with age (CITATION). This is caused by the time delay between the core
% collapse supernova (SNII), in which the majority of $\alpha$ elements
% are produced, and the SNIa that primarily produce iron. The older an
% SSP, there is relatively more iron, bringing $[\alpha/{\rm
%   Fe}]$ to lower values. The Chabrier
% distribution is unimodal and peaks at around a value of $[\alpha/{\rm
%   Fe}]\sim0.5$. The metallicity dependent model is broader and
% possibly bimodal with a second peak at lower values.
%
% The MDF of the gas in the whole halo and in 0.1\,\Rvir is shown in
% figure \ref{fig:gasOoverH} as 12+log(O/H). We see that the gas in the
% IMF(Z) model is
% bimodal over the whole halo, but unimodal in the central galaxy. Since
% there are less supernova at later times, the galaxy does not feed the
% halo with metals. There is less interaction between the galaxy and halo.
%
%% 
\begin{figure*}
  \centering
 \includegraphics[width=0.95\textwidth]{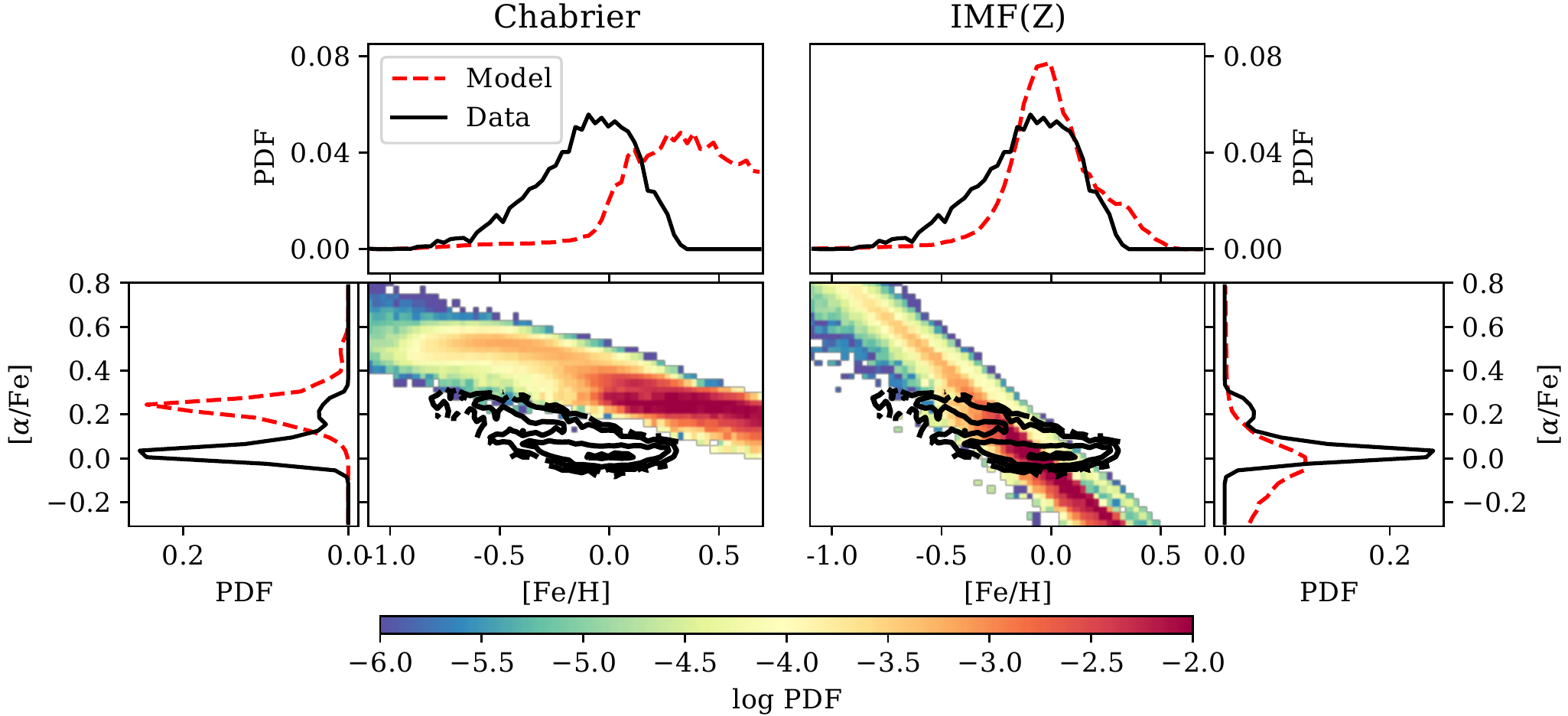}
 \caption{Distribution of simulated stars in the [$\alpha$/Fe] vs
   [Fe/H] plane for Chabrier (left panels) and metallicity-dependent
   IMF (right panels). The black contours are the APOGEE data
   \citep{Bovy2014,Ness2017}. We attempt to mimic the age selection
   function of APOGEE. The spatial selection for both the model and
   APOGEE is the following: $5<R_{\rm xy}/{\rm kpc} <9$ and
   $|z|/{\rm kpc} < 1$. The bin width is set to $0.05$ in abundance
   space for both axes. Shown is the $z=0$ simulation result of Halo
   16 (level 4).}
  \label{fig:apogee}
\end{figure*}
%==========================================================================%
\subsection{$\alpha$-Enhancement}
\label{sec:alpha}
%==========================================================================%
%
\begin{table}
  \centering
  \begin{tabular}{l|c|c|c|c}
\hline
    Model&m&y$_0$&m$_{\rm A}$& y$_{0, {\rm A}}$\\
\hline
Chabrier&-0.21&0.36&-0.15&0.30\\
IMF(Z)&-0.77&0.02&-0.79&-0.09\\
\hline
  \end{tabular}
  \caption{Fit parameters for the [$\alpha$/Fe] -- [Fe/H] relation
    with the linear function [$\alpha$/Fe] =  m $\cdot$ [Fe/H] + y$_0$. The fit
    parameters for the data after the APOGEE selection function are also
    shown, indicated by the index A.}
  \label{tab:fit}
\end{table}
\begin{figure}
  \centering
 \includegraphics[width=0.45\textwidth]{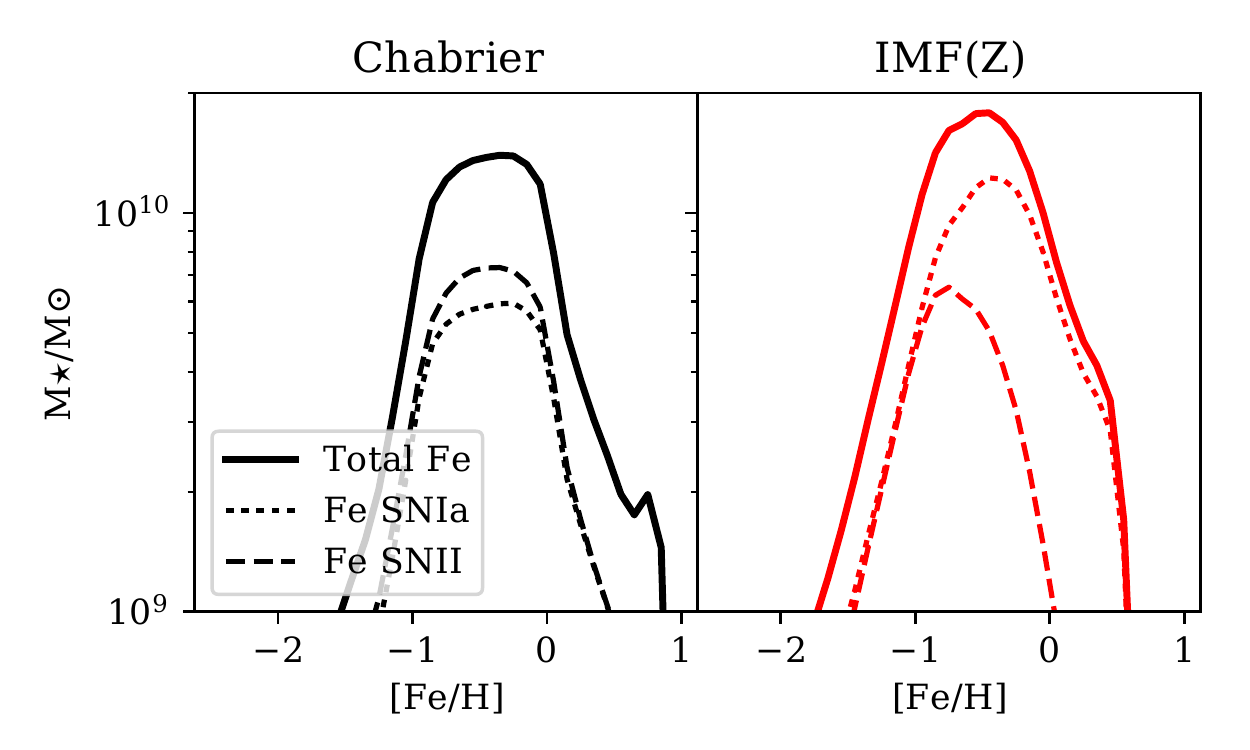}
 \caption{Distributions of [Fe/H] for Chabrier and
   metallicity-dependent IMF. Dotted lines show the iron component
   created in SNIa. Dashed lines show the iron from SNII. Displayed is
   the $z=0$ simulation result of Halo 16 (level-4 resolution).}
  \label{fig:SNFe}
\end{figure}
The so-called $\alpha$-enhancement is an important independent
constraint on simulations. It has often been disregarded in the
comparison with observations since it is difficult to match and
depends on many factors in the simulations. In particular, it depends
not only on the IMF, but also on the prescription for SNIa, the ratio
of SNIa to SNII and, importantly, the yield sets used.

\citet{Marinacci2014} previously explored the $\alpha$-abundance
patterns predicted in {\small AREPO} simulations, finding a need to increase
the SNIa rate by a factor of $\sim 4$ and being prompted to add fully
metal-loaded winds to reproduce the slope of the observations. For a
study of the $\alpha$-abundances of the Auriga sample, see
\citet{Grand2017b}. We show the [$\alpha$/Fe]-[Fe/H] plane in
Figure~\ref{fig:apogee}. Our estimate of $\alpha$ is calculated by
averaging the metallicity contribution of oxygen, silicon and
magnesium.

To compare our simulations with the Milky Way, we show data from the
SDSS Apache Point Observatory Galactic Evolution Experiment (APOGEE),
which is a high-resolution near-infrared spectroscopic survey
\citep[first presented in][]{Bovy2014}. The chemical abundances were
updated to a homogeneous sample by \citet{Ness2017}. We choose stars
in the radial range $5<R_{\rm GC}/{\rm kpc}<9$, where $R_{\rm GC}$ is
the distance to the Galactic center in the plane of the disk. The
scale height is confined to $|z|/{\rm kpc}<1$. APOGEE uses red clump
stars and these have a strong age-dependent selection function,
described in \citet[][their equation 11]{Bovy2014}. We mimic this
selection function when analysing the simulation data, which is
strongly biased towards young stars.

The left panels show the data for Chabrier, and the right ones for the
metallicity dependent IMF. The colored images display the distribution
of the simulation, while the overplotted black contours show the
APOGEE data. The graphs to the top and sides are the one-dimensional
histograms of the same data. Here the simulations are shown as red
dashed lines, while APOGEE is given through black solid lines.  The
variable IMF model causes a significant change in the $\alpha$
enhancement of the stars. The distribution becomes much steeper and
the mean matches the data better. The shape and slope of the
distribution in the Chabrier model seem to match the data more
closely, but in this case the normalization is off by around
0.3~dex. We fit the slope of the [$\alpha$/Fe]-[Fe/H] distribution
with and without the APOGEE selection function. The fit results are
noted in Table~\ref{tab:fit}. In the Chabrier case, the selection
function causes the fit to be shallower, the slope being 0.21 without
and 0.15 with it. This is opposite in the variable IMF case, where the
slope becomes 0.79 with the age selection and 0.77 without.

We note that the bimodality in [$\alpha$/Fe] of the Chabrier model is
likely due to the age-selection, since the total distribution in the
simulation does not show this trend. We assume that the overlapping
functional forms of the SFH (which peaks at $z\sim2$) and the
age-dependent selection function (which peaks around $z\sim0.15$) can
cause an apparent dip in the distribution, making it seem
bimodal. {Our IMF(Z) model does not show a bimodal trend in [$\alpha$/Fe]. As \citet{Grand2017b} have shown, the occurence of the bimodality in simulations is very dependent on the SFH and can also be hidden in some systems. The altered SFH in the variable model may thus well account for the lack of a bimodality. Additionally, it is possible that this result is sensitive to the specific SNIa prescription that is used.}

In Figure~\ref{fig:SNFe}, we show the [Fe/H] distribution of stars in
the entire halo. The dashed and dotted lines are the components of the
distribution split by the channel in which the iron was produced. In
the Chabrier model, SNIa and SNII produce approximately equal amounts
of iron. In the variable IMF model, most of the metal-rich stars have
their iron from SNIa alone. This difference in iron production is the
driving factor that changes the $\alpha$-enhancement between the two
models. The increased iron production at late times (and high
metallicity) brings the [$\alpha$/Fe] ratio down to solar values and
causes the steep dependence in the distribution.
%
%
%==========================================================================%
\section{DISCUSSION}
\label{sec:discussion}
%==========================================================================%
{This study differs significantly from previous work studying a variable
IMF in simulations since it implements the IMF variability {\itshape ab initio} rather
than in post-processing and introduces only a single additional parameter. 
This allows a detailed analysis of the non-linear
effects of both the metallicity enrichment and the amount of
feedback generated by the changing IMF across time.
Additionally, it is a study of how the IMF may vary
within a single galaxy rather than across the galaxy population.

One study that examines the effects of different IMFs on individual galaxies is the work by
\citet{Few2014}. They test three universal IMFs \citep[the models
presented in][]{Salpeter1955, Kroupa1993, Kroupa2001} by running
multiple manifestations of a simulated L* galaxy. They note that more top-heavy IMFs are more efficient at
forming stars and attribute this to the strong replenishment of gas
caused by the larger fraction of high-mass stars. Although this study
does not include a varying IMF, it nevertheless exhibits some of the
same features as our work. At earlier times, when the metallicity is
still low and the IMF still top-heavy in our simulations, the SFR is
also higher due to the large amount of gas injected into the ISM from
the increased number of supernovae. And yet, in our varying model the
SFR drops after this initial burst, since the increased supernovae
drive metal enrichment and, thus, more bottom-heavy IMFs in the
subsequent cycle of star formation.

% However, this fact alone cannot predict the 
% net effect on feedback and chemical abundances. 
% Simulations
% In the context of simulations of
% individual galaxies, \citet{Bekki2013} and \citet{Few2014} also explore
% how a variable IMF can alter a galaxy's chemical evolution.
%
Possibly the most similar study to ours is by \citet{Bekki2013}. They
implement ab initio IMF slopes that are allowed to vary with the physical
properties of star-forming clouds. They run idealized chemo-dynamical simulations of star-forming
galaxies in the mass range $10^{10}-10^{12}\Mhalo$ and vary the
three IMF slopes of the Kroupa IMF in time following the
results presented in \citet{Marks2012}. The low-mass and
intermediate mass slopes vary proportionally to [Fe/H] of the
star-forming gas. The high-mass slope anti-correlates roughly
log-linearly with the density of a high-density gaseous
core where star formation occurs and correlates linearly with [Fe/H].

They find that in strongly star-forming environments, high-density
molecular gas clouds form. These drive the high-mass slope to smaller
values (a more top-heavy IMF), which in turn increases the number of supernovae and the
amount of feedback. Thus, the total SFR decreases with respect to
their universal IMF model. The increased supernova output sets the
final metallicity and $\alpha$-enhancement above fiducial values.

This is similar to the observational results from
\citet{Clauwens2016}, who show that a metallicity-dependent
IMF can speed up the quenching process. They maintain that a lack of feedback at late
times will cause a last burst of star formation that empties the gas
reservoir.
While their study
uses idealized simulations instead of the full
cosmological context our of work,
it is nevertheless instructive that our single parameter model shows opposite trends, where
SFRs are increased in the varying model and total metallcity is
decreased. The most apparent reason for this dissimilar result is that
\citet{Bekki2013} include a second parameter, the density dependence in his high-mass
slope variation. This produces top-heavy IMFs even at later times when
the metallicity has increased and would produce a bottom-heavy IMF in
our model.

Also, \citet{Recchi2015} study the IMF variation proposed by
\citet{Marks2012} in the context of the IGIMF theory. To better match
the mass-metallicity relation they include the SFR as a further
parameter, which determines the exponent of the embedded cluster mass
function. They conclude that any model that allows
the IMF to be more top-heavy in more massive systems will be able to
reproduce the mass-metallicity relation. 

With accumulating observational and theoretical efforts in
understanding possible IMF variations, it is becoming clearer that if
an intrinsic IMF variability exists, its effect is to produce
shallower high-mass slopes (or a more top-heavy IMF) at early times
and evolve to become more bottom-heavy toward $z=0$
\citep{Dave2008}. In our model this trend results directly from the
fact that the total metallicity is a monotonically rising
function with time. However,
although the additional density parameter in \citet{Bekki2013} produces a
non-monotonic change of the IMF slope with time, it follows the
SFH which drops towards $z=0$.
Given the variety of models attempting to study IMF variations, it is
also clear that theoretical work has not yet converged on a consensus.
But since the IMF is a fundamental parameter in many models of star and
galaxy formation, any deviation from a universal shape will have far
reaching consequences. 
}
\section{SUMMARY}
\label{sec:summary}
%==========================================================================%

The strength of energetic feedback from supernova and black holes is
not well constrained by observations. Although simulations require
some amount of feedback to control the run-away character of star
formation and to match empirical relations such as the fundamental
plane and the $\Mstar-\Mvir$ relation, the (sub-grid) parameter choices used in
simulations cannot be determined from first principle calculations
but rather rely on empirical input. Variable IMF models add additional
complications here, because their effects can be degenerate with the
feedback models themselves.  Thus far, theoretical investigations have
largely sidestepped this problem and focussed on constraining feedback
under the assumption of a universal IMF, for simplicity.

In this study, we investigate the effects of a metallicity-dependent
IMF that is constrained by observations of a sample of ETGs from the
CALIFA survey. We have quantified the differences exhibited by Milky
Way analogues whose IMF varies with the total metallicity of the gas
from which each star particle (representing a single stellar
population) forms with respect to ``standard'' galaxies with a
universal Chabrier IMF. We did not recalibrate the feedback strength
or adjust any other parameters in the simulations, thus our results
expose the effect of the variable IMF in isolation. This maximizes the
corresponding effects and thus brackets the uncertainty due to a
possible environmental dependence of the IMF.

One interesting outcome of our variable IMF model is that it produces
too many stars to match the $\Mstar-\Mvir$ relation. This results from
a slight change of the star formation history, generally being higher
than the Chabrier model around $z\sim 2-3$. 
%Our simulations do not show such a trend. Instead,
Late time SFRs
vary between halos. Halos 9 and 28 show a slightly lower SFR compared to the
Chabrier IMF in the
last $\sim2\,$Gyr. Whereas the SFR of Halo 6 does not significantly change and
Halo 24 has $\sim0.5$\,dex higher SFR since $z\sim0.3$ (compared to
the Chabrier IMF halos).

The most significant
result is that the metallicity evolution rises faster and flattens out
early, which produces lower total metallicities. The MDF of stars is
then less broad and more peaked.
We have investigated the metal production further by examining the
$\alpha$-abundance and its dependence on iron abundance. The
corresponding distributions are broader in the variable IMF model, and
even hint at a bimodality. To compare this accurately with
observations, we mimic the selection function of stars in the APOGEE
survey. This includes a spatial selection in the disk and in terms of
the scale height. But since APOGEE observes primarily red clump stars,
it also includes an age selection \citep[see][]{Bovy2014}. The
distribution in the [$\alpha$/Fe]-[Fe/H] plane is strongly influenced
by the IMF model, with the slope of the relation being much steeper in
the variable model. This model in fact matches the APOGEE data
considerably better, since its mean is closer to the observed
values. Although the Chabrier model does not pass through the data,
its slope appears closer the data, but its normalization is too high
by 0.3 dex. Overall, these results highlight the significant
constraining power in detailed abundance patterns for fundamental
questions in galaxy formation such as the degree of universality of
the stellar IMF. More detailed simulation studies will be needed to
fully understand the non-linear cross-talk between feedback modelling,
chemical enrichment, and the variability of the IMF.

% 
% It seems that the slope in the [$\alpha$/Fe]-[Fe/H] plane is an
% indicator of the metallicity dependency of the IMF. If the density
% distribution follows the SFH, and the
% bimodality is caused by two major episodes of star formation in the
% Milky Way's past, the change in slope between the older stars (the
% thick disk component) and the younger ones would indicate a stronger
% metallicity dependency in the past. 
%
%==========================================================================%
\section*{Acknowledgments} 
%==========================================================================%
%
We thank an anonymous referee for suggestions and constructive
criticism which considerably improved the quality of this paper.

The authors also thank Robert Grand, Hans-Walter Rix and R\"{u}diger Pakmor
for useful conversation.  TAG and VS acknowledge funding through the
Collaborative Research Centre SFB 881 ``The Milky Way System''
(subproject A1) of the German Research Foundation (DFG).  This
research was carried out on the High Performance Computing resources
of the {\sc draco} cluster at the Max Planck Computing and Data
Facility (MPCDF) in Garching operated by the Max Planck Society (MPG).
\balance
\bibliographystyle{mnras}
\bibliography{bib}
\appendix
\bsp
\label{lastpage}
\end{document}